\documentclass[showpacs,12pt,preprintnumbers,amsmath,amssymb,aps,nofootinbib]{revtex4}

\newcommand{\step}{\vspace{.5em}}

\def\di{\displaystyle}

\def\bg{\begin{eqnarray}\begin{array}{rcl}\displaystyle}
\def\eg{\end{array} &\di    &\di   \end{eqnarray}}
\def\bm#1{\begin{eqnarray}\begin{array}{#1}\di}
\def\bmo#1{\begin{eqnarray*}\begin{array}{#1}\di}
\def\bml#1#2{\begin{eqnarray}\begin{array}{#1}\label{#2}\di}
\def\bgo{\begin{eqnarray*}\begin{array}{rcl}\displaystyle}
\def\ego{\end{array} &\di    &\di \nonumber  \end{eqnarray*}}

\def\btensor#1#2{\renew\left#1\begin{array}{#2}\di}
\def\brtensor#1#2#3{\ren#3\left#1\begin{array}{#2}}
\def\botensor#1#2{\renew\left#1\begin{array}{#2}}
\def\etensor#1{\end{array}\right#1}

\def\eq#1{(\ref{#1})}
\def\Eq#1{Eq.~(\ref{#1})}

\def\tr{{\rm tr}}
\def\Tr{{\rm Tr}}

\def\ov{\over}
\def\s0#1#2{\mbox{\small{$ \frac{#1}{#2} $}}}
\def\0#1#2{\frac{#1}{#2}}

\def\del{{\mbox{\boldmath$\delta$}}}
\def\del{\delta}

\def\CO{{\mathcal O}}

\def\ren#1{\renewcommand{\arraystretch}{#1}}

\def\renew{\renewcommand{\arraystretch}{1}}

\begin{document}

{\normalsize\begin{flushright}
CERN-PH-TH/2006-183,
SHEP-06-27, 
HD-THEP-06-23\\[10ex] \end{flushright}
}

\title{Non-perturbative thermal flows and resummations}

\author{Daniel F.~Litim${}^{a,b}$ and Jan M. Pawlowski${}^c$}

\affiliation{\mbox{${}^a$ School of Physics and Astronomy, U Southampton,
Highfield, SO17 1BJ, U.K.}\\
\mbox{${}^b$ Physics Department, CERN, Theory Division, CH -- 1211 Geneva 23. 
}\\
 \mbox{${}^c$ Institut f\"ur Theoretische Physik, Universit\"at Heidelberg, }
\\\mbox{Philosophenweg 16, D-69120 Heidelberg, Germany}}%

\begin{abstract}
${}$\\[-1ex]

\centerline{\bf Abstract} We construct a functional renormalisation group for
thermal fluctuations. Thermal resummations are naturally built in, and the
infrared problem of thermal fluctuations is well under control. The viability
of the approach is exemplified for thermal scalar field theories. In gauge
theories the present setting allows for the construction of a gauge-invariant
thermal renormalisation group.
\end{abstract}

\thispagestyle{empty}
\maketitle

\section{Introduction} 

An important goal in thermal field theory is the computation of physical
quantities like the pressure or screening masses of quantum fields. This is of
interest for on-going and future experiments at RHIC, LHC and the FAIR
facility, which are sensitive to the quark-gluon-plasma in the core region of
heavy ion collisions.  Recent RHIC data even suggests that a thermal quark
gluon plasma is strongly coupled, see \cite{StrongQGP}. On the theoretical
side, one observes that standard thermal perturbation theory is plagued by
infrared divergences due to massless excitations, which at least requires
resummations.  Resummed thermal perturbation theory often displays a weak
convergence behaviour even deeply in its domain of validity, see,
e.g.~\cite{Kraemmer:2003gd,Andersen:2004fp,Blaizot:2003iq}.  In gauge
theories, the situation is additionally complicated by the magnetic sector.
In our opinion, this asks for a method able to incorporate both the weakly and
the strongly coupled regimes.  Moreover, such a method would benefit from a
clear separation of quantum fluctuations and thermal effects.\step

In the past decade much progress in the description of quantum
fluctuations has been achieved within the functional renormalisation
group, for reviews see
\cite{Morris:1998da,Litim:1998yn,Litim:1998nf,Berges:2000ew,Polonyi:2001se,%
  Salmhofer:2001tr,Pawlowski:2005xe}. The functional RG, applicable both at
weak and strong coupling, has also been used for the computation of
thermal effects,
e.g.~\cite{Tetradis:1992xd,Liao:1995gt,D'Attanasio:1996zt,%
D'Attanasio:1996fy,Comelli:1997ru, Bergerhoff:1998xx,%
Litim:1998yn,Litim:1998nf,Schaefer:1999em,Andersen:1999dy,Freire:2000sx,%
Tflow2003,Berges:2000ew,Baier:2003ex,Braun:2003ii,Blaizot:2004qa,Braun:2006jd,%
Schaefer:2006ds}.  A direct implementation of thermal fluctuations has been
put forward within the real time formalism
\cite{D'Attanasio:1996zt,D'Attanasio:1996fy,Comelli:1997ru} and the imaginary
time formalism \cite{Litim:1998nf}.\step

In the present paper, we extend and detail the proposal
\cite{Litim:1998nf}. We construct thermal flows that are well-defined
both in the infrared and in the ultraviolet, and provide ideal
starting points for numerical and analytical studies. In scalar
theories, we work out the scheme-independent part of the integrated
flow, which contains the one loop thermally resummed perturbation
theory.  We equally provide thermal flows within a thermal derivative
expansion, and reproduce explicitly the leading order results for the
pressure and the thermal self energy. \step

For gauge theories, and for specific momentum cutoff and gauges, the
thermal flow even respects gauge invariance for arbitrary scale $k$.
This follows from two observations. A mass-like momentum cutoff --
even though it is not applicable for integrating-out quantum
fluctuations \cite{Litim:1998nf} -- is a viable cutoff for thermal
fluctuations. The second observation concerns Wilsonian flows for
gauge theories in axial gauges
\cite{Litim:1998qi,Litim:2002ce}.  Gauge invariance for
physical Green functions is controlled via modified Ward-Takahashi
identities.  In an axial gauge they reduce to the standard
Ward-Takahashi identities for any scale $k$, if a mass-like regulator
is employed.  Although of limited use in the generic case, this
is precisely the missing piece to construct a gauge invariant flow for
thermal fluctuations.\step

The outline of the paper is as follows. In Sect.~\ref{FunctionalFlows}, we
review functional flows in the imaginary time formalism. In Sect.~\ref{QvsT},
we discuss our projection method for thermal fluctuations. Scheme independence
of the full one-loop thermal resummation is worked out in
Sect.~\ref{sec:resum}. Explicit thermal flows within a thermal derivative
expansion and results for the thermal pressure and the self-energy are
obtained in Sect.~\ref{sec:deriv} and Sect.~\ref{Opt}.  The extension to
thermal gauge theories is detailed in Sect.~\ref{sec:gauge}, including a
gauge-invariant thermal flow and results for the thermal mass and pressure.
Our conclusions are contained in Sect.~\ref{Discussion}.

\section{Functional flow equations}\label{FunctionalFlows}

To start with, we discuss the standard Wilsonian approach to quantum field
theories. For simplicity, we focus on a bosonic degree of freedom. A
generalisation to fermions and gauge fields is straightforward. All the
physically relevant information can be obtained from the (regularised)
partition function. It reads
\begin{eqnarray}
\label{Schwingerk} \exp W_{k,T}[J]=\int{\cal
  D}\phi \exp\Big(-S_{k,T}[\phi] +\Tr\, J \phi \Big) 
\end{eqnarray}
In $(d+1)$ dimensions and using the imaginary time formalism at temperature
$T$, the trace stands for \begin{eqnarray} \Tr=T \sum_{n} \ 
  \int\0{d^dq}{(2\pi)^d}\ ,
\end{eqnarray}
and the implicit replacements $q_0\to 2\pi n T$ for bosonic 
fields are understood, $n$ labelling the Matsubara frequencies
and $J$ stands for the corresponding sources.
The term $S_{k,T}[\phi]=S+\Delta S_{k,T}$ contains the (gauge-fixed)
classical action $S[\phi]$ and a quadratic regulator term
$\Delta S_{k,T}[\phi]$, given by 
\begin{eqnarray} 
\Delta S_{k,T}[\phi] = 
\012 \Tr\, \Big[ \phi(-q)\ R_k(q^2)\ \phi(q) \Big] . \label{Rk}
\end{eqnarray} \Eq{Rk} introduces a coarse-graining via the operator
$R_k(q)$. The flow of \eq{Schwingerk} related to an infinitesimal change of
$t=\ln k/\Lambda$ (with $\Lambda$ being some fixed UV scale) is
\begin{equation} \label{flowSchwingerk}
{\frac{\partial}{\partial t} \, e^{ W_{k,T}[J]}}=\012 \int{\cal
  D}\phi\, \Tr\left( \phi\, \frac{\partial R_k}{\partial t} \,
  \phi\right)\, e^{-S_{k,T}[\phi] +\Tr\ J \phi}\ .  
\end{equation}
Performing a Legendre transformation leads to the coarse-grained
effective action $\Gamma_{k,T}[\phi]$, 
\begin{eqnarray} \label{ga}
\Gamma_{k,T}[\phi]=\Tr\, J \phi-W_{k,T}[J]-\Delta S_{k,T}[\phi]-C_{k,T}\,, 
\end{eqnarray}
Note that the constant
$C_k$ is usually not mentioned when one is interested in field
independent quantities. However as we shall see later in the
discussion of the thermal pressure we have to take it into account.
It is straightforward to obtain the flow equation for
$\Gamma_{k,T}[\phi]$ by using \eq{ga}: 
\begin{subequations}\label{eq:flow}
\begin{eqnarray}
\label{eq:flow1} \partial_t\Gamma_{k,T}[\phi]=\frac{1}{2}\Tr\,
\left[G_{k,T}[\phi]\ \frac{\partial R_k}{\partial
    t} \right]-\partial_t C_{k,T}\,, \end{eqnarray} 
with 
\begin{eqnarray}
G_{k,T}[\phi]=
\left(\frac{\delta^2\Gamma_{k,T}[\phi]}{\delta \phi\delta \phi}
  +R_k\right)^{-1}\,,
\end{eqnarray}
\end{subequations} 
denoting the full (field-dependent)
regularised propagator of $\phi$. Let us be more specific about the
regulator function $R_k(q)$. The general requirements are
\begin{itemize}
  
\item[($i$)] it has a non-vanishing limit for $q^2 \to 0$, typically
  $R_k\to k^2$ for bosons, and $R_k\to k$ for fermions.
  This precisely ensures the IR finiteness of the propagator at
  non-vanishing $k$ even for vanishing momentum $q$.
  
\item[($ii$)] it vanishes in the limit $k\to 0$, and for $q^2\gg k^2$.
  The latter condition ensures that large momentum fluctuation 
have efficiently been integrated-out whereas the first condition guarantees 
that any dependence on $R_k$ drops out in the limit ${k\to 0}$.
  
\item[($iii$)] $R_k$ diverges like $\Lambda^2$ for bosons, and
  like $R_k\to \Lambda$ for fermions, when $k\to \infty$ (or
  $k\to \Lambda$ with $\Lambda$ being some UV scale much larger than
  the relevant physical scales). Thus, the saddle point approximation
  to (\ref{Schwingerk}) becomes exact and $\Gamma_{k\to\Lambda}$
  reduces to the (gauge-fixed) classical action $S$.
\end{itemize}
These conditions guarantee that $\Gamma_k$ has the limits
\begin{eqnarray}
  \lim_{k\to \infty}\Gamma_{k,T}[\phi]&=&S[\phi]\label{eq:initial}\\
  \lim_{k\to 0}\Gamma_{k,T}[\phi]&=&\Gamma_T[\phi]\ .\label{eq:final}
\end{eqnarray} 
For any given scale $k$ the main contributions to the running of
$\Gamma_{k,T}$ in \eq{eq:flow} come from momenta about $q^2\approx k^2$, if
the regulator function obeys the conditions ($i$)-($iii$).  This is so because
$\partial_t R_k$ is peaked about $q^2\approx k^2$, and sufficiently suppressed
elsewhere. The physics behind this is that a change of $\Gamma_{k,T}$ due to a
further coarse graining (i.e. the integrating-out of a thin momentum shell
about $k$) is dominated by the fluctuations with momenta about $k$.
Contributions from fluctuations with momenta much smaller/larger than $k$
should be negligible. The flow equation \eq{eq:flow} connects the classical
action $S[\phi]$ with the full quantum effective action $\Gamma_T[\phi]$ at
temperature $T$.  The full quantum effective action is hence defined by the
flow equation and an initial action at some UV scale. Note that the limits
\eq{eq:initial} and \eq{eq:final} are strictly speaking only valid with a
suitable choice of $C_k$.

\section
{Quantum vs.~thermal fluctuations}\label{QvsT} 

At finite temperature the flow \eq{eq:flow} constitutes the
integrating-out of quantum as well as thermal fluctuations.
\Eq{eq:flow} can be projected on thermal fluctuations: instead of
computing the flow for $\Gamma_{k,T}[\phi]$ as in \eq{eq:flow}, we
propose to study the flow for the difference 
\begin{eqnarray}\label{eq:diffG} 
\bar\Gamma_{k,T}[\phi]=\Gamma_{k,T}[\phi]-\Gamma_{k,0}[\phi]\ 
\end{eqnarray}
between the effective action at vanishing and non-vanishing temperature
\cite{Litim:1998nf}. \Eq{eq:diffG} is evaluated for fields periodic in
time.\footnote{For $\Gamma_{k,0}$, this means that Greens functions
  $\Gamma_{k,0}^{(n)}$ are multiplied with periodic fields, and integrated
  over the compact time interval.}
It entails the thermal effects in the effective action as we have effectively
removed the quantum fluctuations.  The flow of \eq{eq:diffG} reads
\begin{eqnarray}
&&\hspace{-.6cm}\partial_t\bar\Gamma_{k,T}[\phi]= \012 
\int\0{d^3q}{(2\pi)^3} \Bigl[T\ \sum_{n}
  G_{k,T}[\phi]\, \partial_t R_k
-\int {d q_0\ov 2\pi}
  \,G_{k,0}[\phi]\,\partial_t R_k\Bigr]
-\partial_t \bar C_{k,T} \ ,
 \label{eq:Tflow}  \end{eqnarray} 
where $\bar C_{k,T}=C_{k,T}- C_{k,0}$. The flow \eq{eq:Tflow} vanishes at
$T=0$, and hence is a thermal flow.  It can be used within a two-step
procedure: first, one computes the flow \eq{eq:flow} at vanishing temperature
within a truncation that is adapted to the zero temperature theory. Then the
result is used as an input for the thermal flow \eq{eq:Tflow}. We emphasise
that the truncation at finite temperature may differ even qualitatively from
that at $T=0$. \step

The thermal flow \eq{eq:Tflow} has another important advantage that is
particularly interesting for its application to gauge theories.  We
can relax the conditions $(i)-(iii)$ on the regulator. The aim is to find
a reliable, but still sufficiently simple and manageable formulation
of the flow equation. The first observation is that the flow equations
\eq{eq:flow} and \eq{eq:Tflow} indeed are simplified for a mass-like
regulator given by \bm{c}\label{R-mass}
\begin{tabular}{ll}
  {bosons:\ }&$R_k(q)\sim k^2$\\
  {fermions:\ }&$R_k(q)\sim k$
\end{tabular}
\eg or variants of it.\footnote{Sometimes it is convenient to multiply
  the mass term with a (momentum-independent) wave function
  renormalisation, $R_k=Z_k\ k^2$ (and analogous for
  fermions).  The same reasoning applies.}  The key characteristic of
$R_k$ in \eq{R-mass} is that it does not depend on
momenta.\step

Formally speaking, \eq{R-mass} is a viable IR regulator in the sense
of condition ($i$), and it allows as well to reach the UV initial
condition, due to condition ($iii$). However, the choice \eq{R-mass}
violates condition ($ii$), which is one of the basic requirements for
a Wilsonian cutoff.  Indeed, the operator $\partial_t R_k$ appearing
in \eq{eq:flow} is neither peaked about $p^2\approx k^2$, {\it
  nor} does it lead to a sufficient suppression of high momentum
modes. The flow equation \eq{eq:flow} would then receive contributions
from the high momentum region for any value of $k$.  An
immediate consequence of this is that an additional UV regularisation
is required, as the flow equation \eq{eq:flow} is no longer
well-defined for large loop momenta. Stated differently, on might say
that a mass term regulator leads to a break-down of the Wilsonian
picture, since it is no longer related to an integrating-out of
momentum degrees of freedom. Rather, it corresponds to a flow within
the space of massive theories. \step

Apart from these more formal objections one should mention that numerical
solutions of \eq{eq:flow} are much more involved than for regulators
satisfying $(ii)$. At every iterative step a $d$-dimensional momentum integral
has to be performed in \eq{eq:flow} over a non-trivial function which is not
strongly peaked about some momentum region. This is, numerically, a quite
tedious problem. For this reason most of the sophisticated numerical
investigations are based on non-local regulator functions (like the sharp
cutoff, exponential or algebraic ones). It has also been observed that
approximate solutions to the flow equation (expansions in powers of the field,
derivative expansions) show a rather poor convergence behaviour, when
\eq{R-mass} is used. In gauge theories, this has been seen from explicit
perturbative computations
\cite{Simionato:1998iz,Simionato:2000ut,Panza:2000tg}. Consequently at $T=0$
it is not advisable to directly resort to a mass-like regulator. \step

While \eq{R-mass} is not viable for quantum fluctuations being present in
\eq{eq:flow}, we will now argue that it is perfectly viable for thermal ones
in \eq{eq:Tflow}.  The main point is that the large momentum fluctuations (not
sufficiently controlled by \eq{R-mass}, introducing UV divergences to the flow
equation) have nothing to do with the heat bath. Therefore, subtracting the
zero temperature quantities will render the flow equation \eq{eq:Tflow} finite
and well-defined, even in the presence of a mass-like regulator $R_k=k^2$.
Then the flow \eq{eq:Tflow} boils down to
 \begin{equation}
\partial_t\bar\Gamma_{k,T}[\phi]= 
k^2\int\0{d^3q}{(2\pi)^3} \Bigl[T\ \sum_{n}
  G_{k,T}[\phi]
-\int {d p_0\ov 2\pi}
  \,G_{k,0}[\phi]\Bigr]
-\partial_t \bar C_{k,T} \ .  
 \label{eq:Tflowmass}
\end{equation}
Here, the momentum-independent regulator $\partial_t R_k\sim R_k$ acts only as
a multiplicative constant because of \eq{R-mass}.  In this case, the
suppression of large momenta does not originate from $\partial_t R_k$,
but from the cancellation between the propagator terms.  For large internal
momenta, the Matsubara sum can be replaced by an integral, thereby cancelling
the $T=0$ contribution.  Therefore, one may read \eq{eq:Tflowmass} as a
Wilsonian flow for thermal fluctuations: At the starting point $k=\Lambda$
($\Lambda$ being some large UV scale) all fluctuations are suppressed and
\eq{eq:Tflowmass} vanishes. For any $k<\Lambda$, the flow of
$\Gamma_{k,T}[\phi]$ would receive contributions for all momenta. In contrast,
the difference $\bar\Gamma_{k,T}[\phi]$ is sensitive only to thermal
fluctuations, which are peaked in the infrared region and naturally decay in
the UV region.  It follows that the integrand in \eq{eq:Tflowmass} is peaked
about $q^2\approx k^2$. In other words, condition $(ii)$ is effectively
guaranteed even in the case of a mass-like regulator by the very nature of the
temperature fluctuations. This amounts to the fact that the mass-like
regulator seems to be a reasonable choice even for numerical applications in
thermal field theories.\step 

It is worth mentioning that the cancellation of UV divergences in
\eq{eq:Tflowmass} is reminiscent of the BPHZ-procedure. There, the
subtraction of possibly divergent terms takes place on the level of the
integrand rather than on the level of the regularised full expressions. 
\step
 
Let us comment on the initial condition to \eq{eq:Tflowmass} [and
\eq{eq:Tflow}].  In contrast to the flow \eq{eq:flow} with the limits
\eq{eq:initial} and \eq{eq:final}, the flow equation \eq{eq:Tflowmass}
[resp. \eq{eq:Tflow}] has the limits 
\begin{subequations}\label{eq:bound}
\begin{eqnarray}
\lim_{k\to\infty} \bar \Gamma_{k,T}[\phi]&=&0\label{eq:diff-initial}\\
\lim_{k\to 0} \bar \Gamma_{k,T}[\phi]&=&\Gamma_T[\phi]-
\Gamma_0[\phi]\label{eq:diff-final}\ .  
\end{eqnarray} 
\end{subequations}
The boundary condition \eq{eq:diff-initial} looks rather simple. The flow
equation \eq{eq:Tflowmass} needs in addition the knowledge of the massive
$T=0$ quantum theories.  This point is qualitatively shared by the proposals
\cite{D'Attanasio:1996zt,Drummond:1996ap}. It seems likely to find a good
approximation for the issues under investigation, since \eq{eq:Tflowmass} is
eventually projecting-out thermal fluctuations. Those should not be too
sensitive to the details of the quantum effective action at $T=0$.  Moreover
we deal with a situation were the original fields are still sensible degrees
of freedom. Thus, a perturbatively resummed quantum effective action should be
a good starting point.  Here we are actually taking advantage of the fact that
for a mass-like regulator the flow is describing a path in the set of massive
vector boson theories rather than a Wilsonian integrating-out.

\section{Thermal resummations} \label{sec:resum}
 
The present formulation naturally incorporates thermal resummations
within general truncation schemes. Moreover, an important, direct,
consequence of the flow \eq{eq:Tflow} is scheme independence of lowest
order resummed perturbation theory: at this order the effective action
at $k=0$ does neither depend on the regulator nor on the truncation.
This is proven below, and put to work within the derivative expansion
in the next section. We discuss a scalar $\phi^4$-theory with a
massless (at $T=0$), neutral scalar field and coupling $\lambda$,
\begin{eqnarray}\label{eq:Scl}
S_{\rm cl,T}[\phi]=-\012 \int_x \phi(x)\,\partial^2\phi(x) 
+ \0{\lambda}{4!}\int_x \phi^4(x)\,, 
\end{eqnarray}
where $\int_x=\int_0^{1/T} d\tau \int d^3 x$. We are interested in the
$\lambda$-dependence of thermal corrections. It is well-known that
naive perturbation theory breaks down beyond one-loop, and we are left
with an expansion in $\lambda^{1/2}$ rather than in $\lambda$. For the
proof of scheme independence we rewrite the flow \eq{eq:Tflow} as
\begin{equation}
\partial_t \bar\Gamma_{k,T}= \012 
  \Tr 
  \left[\0{1}{\Gamma^{(2)}_{k,T}+R_k}-\0{1}{\Gamma^{(2)}_{0,T}+R_k}
  \right] \partial_t R_k+\012 \Tr \0{1}{\Gamma^{(2)}_{0,T}+R_k}\partial_t 
R_k-\012\Tr \0{1}{\Gamma^{(2)}_{k,0}+R_k}\partial_t R_k
\label{eq:Tflows} 
\end{equation}
where we have dropped the constant contribution. The traces $\Tr$ in
\eq{eq:Tflows} have to be taken at temperature $T$ (first terms), and
$T=0$ (last term). For computing the one loop effective action from
\eq{eq:Tflows} we have to insert the classical action on the rhs of
\eq{eq:Tflows}. Then the first term on the rhs of \eq{eq:Tflows}
vanishes. The remaining terms are total $t$-derivatives and can be
integrated trivially. After a reordering we end up with
\begin{equation}
 \Gamma^{{\rm 1-loop}}_{k,T}=\left[
\012 \Tr\ln \0{S_{\rm cl,T}^{(2)}+R_{k}}{ S_{\rm cl,T}^{(2)}+R_{\Lambda}}
+\Gamma^{{\rm 1-loop}}_{\Lambda,T}\right]
  -\left[\012 \Tr\ln \0{S_{\rm cl,0}^{(2)}+R_{k}}{S_{\rm cl,0}^{(2)}
+R_{\Lambda}}
-(\Gamma^{{\rm 1-loop}}_{k,0}-\Gamma^{{\rm 1-loop}}_{\Lambda,0})\right]\,.
\label{eq:1loop}\end{equation} 
The second bracket in \eq{eq:1loop} vanishes identically due to
\eq{eq:flow}.  The $\Lambda,T$-dependent terms in \eq{eq:1loop} just
arrange for the appropriate renormalisation; indeed their sum has to be 
$\Lambda$-independent. This results in 
\begin{eqnarray}\label{eq:G1loop} 
  \Gamma^{\rm 1-loop}_{k,T}=\012 \int_x \phi(x)\,\left(\partial^2
+\Pi_{k,T}^{\rm 1-loop}\right)\phi(x) 
  + \0{\lambda}{4!}\int_x \phi^4(x)+\CO(\lambda^2)\,,  
\end{eqnarray}
where $\Pi_{k,T}^{\rm 1-loop}=m_{T}^2(1+\CO(k/T))$ is the
momentum-independent self-energy at one loop, and $m_{T}^2\propto
\lambda$ is the standard thermal mass at $k=0$. Its scheme
independence follows from \eq{eq:1loop}. At $k=0$ we are left with the
renormalised one-loop effective action at finite temperature
$\Gamma^{\rm 1-loop}_{0,T} =\Gamma^{\rm 1-loop}_{0,0} +\Delta
\Gamma^{\rm 1-loop}_{0,T}$. Now we re-insert $\Gamma^{\rm
  1-loop}_{k,T}$, \eq{eq:G1loop} on the rhs of \eq{eq:Tflows}, and
project the flow \eq{eq:Tflows} on the leading resummed term of order
$\lambda^{3/2}$. This term only receives contributions from the
vanishing Matsubara frequency $n=0$. Due to the subtraction the first
term on the rhs of \eq{eq:Tflows} has an additional (dimensionless)
power of $\lambda\, k/T$. If integrating the zero frequency flow from
$\Lambda\to\infty$ to $k=0$, we can rescale all expressions with the
one loop self energy $\Pi_{0,T}=m^2_T$. The first term in
\eq{eq:Tflows} is of order $\lambda^2$ due to the additional power
$\lambda k/m_T\propto \lambda^{1/2}$, whereas the second term receives
$\lambda^{3/2} T^2$-contributions from the thermal trace at $k=0$.
This term is a total $t$-derivative and we arrive at
\begin{eqnarray}\label{eq:3/2}
\Gamma_{0,T}=\012 \Tr\ln \0{\Gamma^{(2)}_{0,T}}{
\Gamma^{(2)}_{0,T}+R_{\Lambda}}+\Gamma_{\Lambda,T}
+{\cal O}(\lambda^2)\,, 
\end{eqnarray}
where the $O(\lambda^2)$-terms also include additional subtractions.
\Eq{eq:3/2} includes the full 1-loop resummed perturbation theory, but
also applies to more general truncations of $\Gamma_{0,T}$. We
emphasise that \eq{eq:3/2} is valid for general regulators and truncations and
hence is scheme independent. Due to the scheme independence of the one loop
effective action $\Gamma_{0,T}$, \eq{eq:3/2} scheme-independently contains the
full bubble resummation that leads to the lowest order in the thermal
resummation.  We conclude that the flow \eq{eq:Tflow} leads to the full
thermal resummation at lowest order for any regulator and any truncation.
This includes in particular the thermal mass contribution at order
$\lambda^{3/2}$ as well as the pressure contributions at order $\lambda$ and
$\lambda^{3/2}$. In scalar theories this is a direct consequence of the fact
that these terms are only related to bubble diagrams.

\section{Thermal derivative expansion}\label{sec:deriv} It is useful
to discuss the above results in a very simple truncation, the leading
order in the derivative expansion at finite temperature. In this
approximation, the effective action for a real scalar field $\phi$
reads
\begin{eqnarray}\label{eq:Geff} 
  \Gamma_{k,T}[\phi]=\012 \Tr \phi(-p) p^2 \phi(p)+ \int_x V_{k,T}(\phi)\,, 
\end{eqnarray} 
with the effective potential $V_{k,T}(\phi)$. We follow the reasoning of
Sect.~\ref{QvsT} and consider the flow for the thermal difference
$U_{k,T}(\phi)=V_{k,T}(\phi)-V_{k,0}(\phi)$, explicitly given by
\begin{equation}
\partial_t U_{k,T}= \012 
\int\0{d^3q}{(2\pi)^3} 
\left[T\ \sum_{n}
\frac{\partial_t R_k}{(2\pi n T)^2+{\bf q}^2+R_k+V''_{k,T}}
-\int {d q_0\ov 2\pi}
\frac{\partial_t R_k}{q^2+R_k+V''_{k,0}} \right]\,.
 \label{eq:TflowPot}  
\end{equation} 
We have dropped a field-independent constant, and $q^2=q_0^2+{\bf q}^2$.
For lowest order resummed perturbation theory we can concentrate on the 
vanishing Matsubara frequency. The flow \eq{eq:TflowPot} then reads 
\begin{equation}
\partial_t U_{k,T}= \012 T 
\int\0{d^3q}{(2\pi)^3}
\frac{\partial_t R_k}{{\bf q}^2+R_k+V''_{k,T}}
+\CO(\lambda^2)= \012 T 
\int\0{d^3q}{(2\pi)^3}
\frac{\partial_t R_k}{{\bf q}^2+R_k+V''_{0,T}}
+\CO(\lambda^2)\,, 
 \label{eq:TflowPot0}  
\end{equation} 
and \eq{eq:3/2} follows.\step 

At finite temperature a regularisation of the
theory is already achieved for cutoffs depending only on spatial
momenta, $R_k=R_k({\bf q})$. These cutoffs are well-adapted to the
present situation: then the thermal flows \eq{eq:Tflow} successively
sum up thermal fluctuation at momenta ${\bf q}^2\approx k^2$, and the 
thermal properties of the theory are unchanged. Technically this has 
the benefit that the Matsubara sum is unchanged. In the present case 
this allows us to perform the Matsubara
sum and the $q_0$-integration in \eq{eq:TflowPot} analytically, 
\begin{equation}\label{eq:flowspat} 
\partial_t U_{k,T}
=\012 \int \0{d^3 q}{(2 \pi)^3} \partial_t R({\bf q})\left[
\012 \left(\0{1}{\omega_{k,T}}-\0{1}{\omega_{k,0}}\right)
+\0{ n(\omega_{k,T})}{\omega_{k,T}}\right]\,, 
\end{equation}
where $\omega_{k,T}=\left({\bf q^2} + R_k({\bf
    q})+V''_{k,T}\right)^{1/2}$. 
We note the appearance of the Bose-Einstein distribution 
\begin{eqnarray}\label{eq:omegan}
\quad n(\omega)=\0{1}{
e^{\omega/T}-1}
\end{eqnarray} 
in the flow, a direct consequence of the thermal sum.  Every single
term on the rhs in \eq{eq:flowspat} remains well-defined even for large
spatial momenta, due to $\partial_tR$. Along the flow, \eq{eq:flowspat} uses
the zero temperature running for $V_{k,0}$.  While the last term in
\eq{eq:flowspat} is clearly of a thermal origin, the first two terms display a
non-trivial cancellation of zero temperature and thermal contributions. \step

In the high temperature limit, the Bose-Einstein distribution is
strongly enhanced, with $n(\omega)\to T/\omega$. Consequently, the
flow \eq{eq:flowspat} is parametrically dominated by the last term for
$T\to \infty$, and fixed $k$. In this limit, we introduce rescaled
fields as $\varphi= \phi/\sqrt{T}$ and the rescaled potential as
$V_k(\varphi)=U_k(\phi)/T$. Then
\begin{equation}\label{eq:flowHighT} 
\partial_t V_k
=\012 \int \0{d^3 q}{(2 \pi)^3} 
\0{\partial_t R({\bf q})}{{\bf q^2} + R_k({\bf q})+V''_{k}}\,.
\end{equation}
All temperature dependence has disappeared and we are left with a
standard flow in three dimensions. We emphasise that \eq{eq:flowHighT}
equals the zero mode flow in \eq{eq:TflowPot0}, as all higher modes
decouple due to their masses $2 \pi T\to \infty$. We also point out
that \eq{eq:flowHighT}, unlike \eq{eq:flowspat}, is closed since
contributions from $\omega_{k,0}$ are absent in this limit.\step

In the low temperature limit $T\to 0$, the thermal flow \eq{eq:flowspat}
vanishes identically. At finite temperature, and in a one-loop approximation,
$V_{k,T}$ is substituted by the classical potential $V_{\rm cl}$ and the first
two terms in \eq{eq:flowspat} vanish.  Then the thermal corrections reduce to
a total $t$-derivative, and we obtain the cutoff-independent result
\begin{equation}\label{1-loop}
V^{\rm 1-loop}_{0,T}=V^{\rm 1-loop}_{0,0}+
T\int\0{d^3q}{(2\pi)^3}
\ln\left[1-\exp(-\sqrt{{\bf q^2}+V''_{\rm cl}}/T)\right]
\end{equation}
upon integration. This equals \eq{eq:1loop} in the present truncation.
From \eq{1-loop}, we derive the free pressure for a single bosonic
degree of freedom $P[T]=\pi^2T^4/90$ (where $P=-V$). 
At this order, the self-energy correction to the propagator
$\Gamma^{(2)}_{k,T}({\bf q}) = \Pi_{k,T}^{{\rm 1-loop}}({\bf q})$ is
momentum-independent and given by a thermal mass $\Pi_{0,T}({\bf
  q^2}=0)=m^2_{T}=\lambda T^2/24$. We have proven in
Sect.~\ref{sec:resum} that the next-to-leading order of the
self-energy is also scheme independent.  This is most directly seen by
studying \eq{eq:TflowPot0} for the self-energy.  The $n=0$ Matsubara
mode generates a $\lambda^{3/2}$-term from the leading order thermal
mass.  The scheme independent coefficient follows from integrating the
total $t$-derivative in \eq{eq:TflowPot0}, leading to $\delta
m^2_T=-\lambda\, T\, m_T/(8\pi)$.  Hence,
\begin{eqnarray}\label{eq:3/2Pi}
  \Pi_ {0,T}=\0{\lambda T^2}{24}
\left[1-3 \left(\0{\lambda}{24\pi^2}\right)^{1/2} +{\cal O}(\lambda)\right]\,.
\end{eqnarray} 
The thermal pressure can be obtained by re-inserting the self-energy
$\Pi_{k,T}$ into the flow \eq{eq:TflowPot}: $\Gamma^2_{k,T}[\phi=0]=
(2\pi n T)^2 +{\bf q}^2+\Pi_{k,T}$. This produces $k$-dependent bubble
diagrams with regulator insertions. The lowest order (two loop vacuum
bubbles) are total $t$-derivatives, and the field-independent part
$\Gamma_{0,T}[0]$ reads
\begin{eqnarray}
&& \hspace{-.6cm} 
P=\0{\pi^2 T^4}{90}\left[1-\0{15}{8} \left(\0{\lambda}{24\pi^2}\right) 
+\0{15}{2}\left(\0{\lambda}{24\pi^2}\right)^{3/2}
+{\cal O}(\lambda^2)\right]\,.
\label{eq:Pscalar}\end{eqnarray} 
In the above discussion, we have focused on scheme independent
aspects of thermal flows. Now we specify explicit momentum cutoffs,
which should be useful for numerical or analytical studies beyond the
present order. Following the reasoning of Sect.~\ref{QvsT}, we
consider first a mass term momentum cutoff $R_k({\bf q})=k^2$. Then
\eq{eq:flowspat} turns into
\begin{equation}\label{LPA:mass}
\partial_t U_{k,T}
=
\frac{k^2}{2}\int\frac{d^3q}{(2\pi)^3}\left[
\frac{1}{\omega_{k,T}}\coth\frac{\omega_{k,T}}{2T}
-\frac{1}{\omega_{k,0}}\right]\,,
\end{equation}%
where $\omega_{k,T}=({\bf q^2}+k^2+V''_{k,T})^{1/2}$ depends on
spatial momenta ${\bf q}$.  At fixed momentum cutoff $k$, we note that
both terms in the integrand diverge quadratically in the UV, while the
sum remains well-defined even for large loop momenta. Furthermore,
terms sensitive to $V''_{k,T}$, $V''_{k,0}$ or their difference are
suppressed by additional powers in $1/{\bf q^2}$ and are, hence,
finite. This is an explicit example of the cancellation emphasised
above.
\step

\section{Optimised thermal flows}\label{Opt}

The physical content of a given truncation and its stability and convergence
towards the physical theory is enhanced by suitably optimised choices for the
Wilsonian cutoff \cite{Litim:2000ci, Litim:2005us,Pawlowski:2005xe}.
Here, the
thermal flows \eq{eq:flowspat} and \eq{LPA:mass} still require a spatial
momentum integration due to the non-trivial ${\bf q}$-dependence of the
integrand.  We take advantage of a simple optimised momentum cutoff 
\cite{Litim:2001up} 
\begin{eqnarray}\label{eq:optspat} 
  R_k({\bf q})=(k^2-{\bf q^2})\theta(k^2-{\bf q^2})\,,
\end{eqnarray} 
which is known to display remarkable stability properties. This regulator cuts
off the propagating momentum modes in \eq{eq:flowspat} homogeneously, since
${\bf q^2}+R({\bf q^2})= k^2$ becomes momentum-independent for all infrared
modes with ${\bf q^2}<k^2$. Furthermore, the remaining loop integration in
\eq{eq:flowspat} is performed analytically, which is a crucial advantage in
view of numerical studies. The result is
\begin{equation}\label{eq:flowopt} 
  \partial_t U_{k,T}=\0{ k^4 }{6\pi^2}  \left[
    \012 \left(\0{k}{\omega_{k,T}}-\0{k}{\omega_{k,0}}\right)
    +\0{k}{\omega_{k,T}} n(\omega_{k,T})\right]\,, 
\end{equation}
where $\omega_{k,T}=\left(k^2+V''_{k,T}\right)^{1/2}$. The truncated
flow \eq{eq:flowopt} clearly displays the thermal flow structure: the
last term is proportional to the thermal distribution $n(\omega)$
which vanishes for $T\to 0$, and requires no UV renormalisation.  The
first term has the subtraction at $T=0$ which removes the quantum
fluctuation. It is reminiscent of the UV-subtraction in thermal
perturbation theory. Here, however, the subtraction is not necessary
to get a finite flow. Without the subtraction, the flow
\eq{eq:flowopt} has been provided as a proper-time flow in
\cite{Braun:2003ii}. As shown in \cite{Litim:2002xm}, such a
proper-time flow derives from \eq{eq:optspat} to leading order in the
derivative expansion.  Beyond this order, it carries inherent
approximations discussed in \cite{Litim:2002xm}.\step

Within the flow \eq{eq:flowopt} the study of the scheme independent
leading order is most convenient due to its analytic structure. We
solve \eq{eq:flowopt} iteratively, starting at
$U_{\Lambda,T}=V_{\Lambda,T}-V_{\Lambda,0}=0$ and vanishing mass at
$T=0$. To leading order in $\lambda$, corrections in $U_{k,T}$ and its
derivatives originate solely from the thermal (last) term in
\eq{eq:flowopt}.  Performing also the $k$-integration we find the free
pressure $P=\pi^2T^4/90$. For the running thermal mass, we find
\begin{equation}\label{FlowMass}
\partial_t m^2_{k,T}=
\frac{k^5}{12\pi^2}
\left[
\left(\frac{n'(\omega_{k,T})}{\omega^2_{k,T}}-
\frac{n(\omega_{k,T})}{\omega^3_{k,T}}
\right)
\lambda_{k,T}
-\frac{1}{2}\left(
\frac{\lambda_{k,T}}{\omega^3_{k,T}}
-\frac{\lambda_{k,0}}{\omega^3_{k,0}}
\right)
\right]\,.
\end{equation}
To leading order in the coupling, we have
$\lambda_{k,T}=\lambda_{k,0}=\lambda$, and only the first term in
\eq{FlowMass} contributes. To this order, the running thermal mass
reads
\begin{eqnarray}\label{eq:1loopPi}
m^2_{k,{\rm 1-loop}}
=\0{\lambda}{12\pi^2} 
\int_k^\Lambda dx\left[ x\, n(x)-x^2\,n'(x)\right]
+m^2_{\Lambda,\rm 1-loop}\,. 
\end{eqnarray}   
The integral can be performed in terms of poly-logarithms. For
$\Lambda\to\infty$ we have $m^2_{\Lambda,T}\to 0$, and at $k=0$ we
find $m^2_{T}=m^2_{0,{\rm 1-loop}}={\lambda T^2}/{24}$.  While the
$k$-running in \eq{eq:1loopPi} is scheme-dependent, the endpoint is
not, and coincides with the standard expression for the one-loop
thermal correction. Beyond the leading order, the running thermal
masses have to be taken into account. In lowest order resummed
perturbation theory the Matsubara zero mode leads to a
$\lambda^{3/2}$-correction to the self-energy. This term is produced
in the flow from scales
\begin{eqnarray}\label{eq:IRlimit} 
k\ll 2 \pi T\,,
\end{eqnarray}  
also assuming $\lambda\ll 1$. In this region the flow effectively
reduces to the high temperature limit \eq{eq:flowHighT} with
$n(\omega)\to T/\omega$, as $T$ is larger than all other scales.  We
derive from \eq{eq:flowopt}
\begin{equation}\label{eq:flowoptIR} 
  \partial_t U_{k,T}=\0{ k^4 }{6\pi^2} \left(
\0{k T}{\omega_{k,T}^2}+\CO(k/\omega_{k,T})\right)\,,   
\end{equation}
leading to 
\begin{equation}\label{eq:FlowMassIR}
\partial_t m^2_{k,T}=-
\lambda_{k,T}\frac{k^4}{6\pi^2}\left(
\frac{kT}{\omega^4_{k,T}}+\CO(k/\omega^3_{k,T})\right)\,.
\end{equation}
for the mass. At scales \eq{eq:IRlimit} the running of the one loop
mass is sub-leading and we can substitute $m^2_{k,\rm 1-loop}\to
m^2_T$ in \eq{eq:flowoptIR},\eq{eq:FlowMassIR}. Moreover we are only
interested in the $\lambda^{3/2}$-term $m^2_{3/2}$. For dimensional
reasons this term is proportional to $\lambda_{k,T}\, T\, m_T=\lambda
\,T\, m_T+{\cal O}(\lambda^2)$.  After rescaling $k$ with $m_T$ we
arrive at the integrated flow
\begin{equation}\label{eq:FlowMassIRint}
  m^2_{3/2}= \lambda 
  \0{ T m_T }{6\pi^2} \int_0^\infty dk\,
  \left(\frac{k^4}{(k^2+1)^2}-1\right)
  =-\lambda \0{T m_T }{8\pi} \,, 
\end{equation}
which leads to \eq{eq:3/2Pi}. The subtraction of $1$ removes the
infrared one loop contribution. We proceed with the
$\lambda^{3/2}$-contribution $P_{3/2}$ to the thermal pressure $P$,
which follows directly from \eq{eq:flowoptIR},
\begin{eqnarray}\label{eq:3/2press} 
  P_{3/2}= 
  \0{ T m_T^3}{6\pi^2} \int_0^\infty dk\,
  \left(\frac{k^4}{(k^2+1)}-k^2+1\right)
  =\0{T m_T^3 }{12\pi} \,. 
\end{eqnarray} 
The subtraction of $k^2-1$ removes the tree level and one loop
infrared contributions. \Eq{eq:3/2press} provides the
$\lambda^{3/2}$-order in \eq{eq:Pscalar}. \step

In Sect.~\ref{sec:resum} we have shown that the results
\eq{eq:FlowMassIRint}, \eq{eq:3/2press} are scheme-independent.
Therefore it should be possible to turn the corresponding flows into
total $t$-derivatives.  We exemplify this structure with an
independent derivation of the thermal pressure up to order
$\lambda^{3/2}$. The flow $\partial_t P_{k,T}$ of the pressure is
given by (minus) \eq{eq:flowopt} at vanishing field. It depends on the
running mass via
$\omega_{k,T}(\phi)=(k^2+m^2_{k,T}+\s012\lambda\phi^2)^{1/2}$ at
$\phi=0$.  Note that we keep a classical $\lambda$ for technical
reasons. The pressure does not depend on this choice. Now we rescale
the mass in $\omega_{k,T}$ with a parameter $\alpha$, $m^2_{k,T}\to
\alpha m^2_{k,T}$. With this modification we get 
\begin{eqnarray}\label{eq:alphapress}
P(\alpha)= P_{\rm free}+\alpha P_1+\alpha^{3/2} P_{3/2}+\CO(\lambda^2)\,, 
\end{eqnarray} 
since $m_{3/2}^2$ does not contribute to $P_{3/2}$, and the
$\lambda$-dependence only originates in $m_{k,\rm 1-loop}$. Now we
take the $\alpha$-derivative,
\begin{eqnarray}\label{eq:m-der} 
  \partial_{\alpha}\, \partial_t P_{k,T} 
  = -m^2_{k,T} 
  \01{\lambda} \partial_{\phi}^2\, \partial_t U_{k,T}|_{\phi=0}
  +\CO(\lambda^2)
  = -\0{1}{\lambda} \, m^2_{k,T}\, \partial_t m^2_{k,T}
  +\CO(\lambda^2)\,,
\end{eqnarray}
where we have used that $\partial_{\alpha}\omega_{k,T} = \lambda^{-1}\
m^2_{k,T} \partial^2_\phi\,\omega_{k,T}$ at vanishing field. Note also
that $\partial_t m^2_{k,T}$ depends on $\lambda_{k,T}-\lambda$ only at
the order ${\cal O}(\lambda^2)$, as already indicated below
\eq{eq:FlowMassIR}.  Taking derivatives as in \eq{eq:m-der} is a
standard computational trick used in thermal perturbation theory where
mass-derivatives help to analytically perform thermal sums and
momentum integrals. Eq.~\eq{eq:m-der} carries an $\alpha$-dependence
since $\partial_t m^2_{k,T}(\alpha)$ is a function of $\alpha$ via
$\omega_{k,T}$. At $\alpha=1$ the flow \eq{eq:m-der} can be expressed
in terms of a total $t$-derivative,
\begin{eqnarray}\label{eq:total}
m^2_{k,T} \partial_t m^2_{k,T}=\012 \partial_t (m^2_{k,T})^2\,.
\end{eqnarray}
\Eq{eq:total} relates to the $t$-derivative of the two loop vacuum
bubble, and contains the full contribution to the pressure of order
$\lambda$. This reduction leads to the diagrammatic structure of
resummed perturbation theory: the flow can be written as a
$t$-derivative of the infrared regularised resummed diagrams.  Now we
restitute the $\alpha$-dependence in \eq{eq:total} with the help of
\eq{eq:alphapress}. By taking the $\alpha$-derivative of
\eq{eq:alphapress} we conclude that 
the one loop term in \eq{eq:total} qis $\alpha$-independent, and the
resummed contributions of order $\lambda^{3/2}$ are proportional to
$\alpha^{1/2}$. Performing the $t$-integration we arrive at
\begin{eqnarray}\label{eq:expand1}
  \partial_{\alpha} P_{0,T}= -\0{1}{2 \lambda} m^4_{T}+
  \0{T}{8\pi} \alpha^{1/2}
  m^3_{T}+\CO(\lambda^2)
  = -\0{T^2\,m^2_T}{48}\left(1-\0{6}{\pi}\alpha^{1/2}
 \0{m_{T}}{T}\right)+\CO(\lambda^2)\,,
\end{eqnarray} 
where we have used \eq{eq:3/2Pi} for substituting the
$\lambda$-dependence with the appropriate powers of $m_{T}$.  We
integrate \eq {eq:expand1} over $\alpha$ from $0$ to $1$, and arrive
at
\begin{eqnarray}\label{eq:expand2}
P_{0,T}= \0{\pi^2 T^4}{90}-\0{T^2}{48}\left(m_{T}^2 -\0{4}{\pi} 
\0{m^3_{T}}{T}\right)+\CO(\lambda^2) \,,  
\end{eqnarray}
where we have used that $P_{0,T}|_{\alpha=0}$ is the tree-level 
pressure. \Eq{eq:expand2} agrees with \eq{eq:Pscalar}. \step

\section
{Thermal gauge theories}\label{sec:gauge}

We now turn to thermal flows for gauge theories. We aim at deriving
well-defined flows for gauge theories in terms of the fundamental degrees of
freedom, the gluons. The inclusion of matter fields is discussed elsewhere.
For specific regulators and gauges this flow is gauge invariant. The classical
Yang-Mills action is given by
\begin{eqnarray}\label{eq:gauge}
S_A[A]=\frac{1}{2} \int_T d^4 x\, \tr\, F^2(A) \,, 
\end{eqnarray} 
with field strength $F_{\mu\nu}=\partial_\mu A_\nu -\partial_\mu A_\nu
+g [A_\mu,A_\nu]$, and $\tr\, t^a t^b=-\delta^{ab}/2$ in the fundamental
representation. The regulator term for gauge fields is given by
 \begin{equation}\label{eq:A-cutoff}
\Delta S_k[A]=
\frac{1}{2}T\sum_n\int\frac{d^3 q}{(2\pi)^4}
A_\mu^a(-q)\,R^{\mu\nu}_{k,ab}(q)\,A^b_\nu(q)
\end{equation}
where the implicit replacement $q_0\to 2\pi n T$ is understood. With the
regularisation \eq{eq:A-cutoff} and using the definition \eq{eq:diffG}, the
flow \eq{eq:Tflow} reads
\begin{eqnarray}
&&\hspace{-.6cm}\partial_t\bar\Gamma_{k,T}[A]= \012 
\int\0{d^3q}{(2\pi)^3} \left[T\ \sum_{n}
  \,G_{k,T}{}_{\mu\nu}^{ab}[A]\, \partial_t R^{\nu\mu}_{k,ba}
-\int {d q_0\ov 2\pi}
 \, G_{k,T}{}_{\mu\nu}^{ab}[A]\, \partial_t R^{\nu\mu}_{k,ba}\right]
-\partial_t \bar C_{k,T} \ ,
\label{eq:Tgaugeflow}  \end{eqnarray} 
In the presence of matter fields $\phi$ and corresponding regulators 
$R_\phi$ additional terms proportional to $\partial_t R_\phi$ will be 
present in \eq{eq:Tgaugeflow}. For mass-like regulators the flow 
\eq{eq:Tgaugeflow} reduces to 
\begin{eqnarray}
\partial_t\bar\Gamma_{k,T}[A]= k^2 \int\0{d^3q}{(2\pi)^3}\, \tr\,\left[
T\,\sum_{n} G_{k,T}[A]\, \partial_t R_{k}
-\int {d q_0\ov 2\pi}\, G_{k,T}[A]\, \partial_t R_{k}\right]
-\partial_t \bar C_{k,T} \ ,
\label{eq:Tgaugeflowmass}  \end{eqnarray} 
where the trace $\tr$ sums over Lorentz and gauge group indices.  So far
we have not specified the gauge fixing procedure. In general, gauge fixing
leads to additional ghost fields $C$ as well as a regularisations $R_C$. Since
the thermal bath singles out a preferred rest frame, we have an additional
Lorentz vector $n_\mu$ at our disposal. Therefore, it is natural to employ an
axial gauge fixing.  An axial gauge has the further advantage that ghost
fields decouple completely from the theory, and possible Gribov copies are
absent. We employ
\begin{eqnarray} \label{eq:axial} 
  S_{\rm gf}[A]=\01{2}
  T \sum_n\int\frac{d^dq}{(2\pi)^4} \ n_\mu A^a_\mu\ \01{\xi n^2}\ n_\nu
  A^a_\nu\, .
\end{eqnarray} 
In \cite{Litim:1998qi,Litim:1998nf,Litim:2002ce} we
discussed the various aspects of an axial gauge fixing \eq{eq:axial}.
In particular we showed that the spurious propagator singularities of
perturbation theory are naturally absent in a Wilsonian approach.
Furthermore, the gauge fixing parameter $\xi$ with mass dimension $-2$
has a non-perturbative fixed point at $\xi=0$. This singles out the
$nA=0$ gauges and tremendously simplifies the problem of gauge
invariance, because it allows for a momentum independent choice of
$\xi$.\step

Gauge invariance for physical Green functions corresponds to the
requirement of a modified Ward Identity (mWI) to hold. Gauge
transformations on the fields are generated by $\del_\alpha$ with action 
\begin{eqnarray}\label{eq:del}
\del_\alpha A= D(A)\alpha\,.
\end{eqnarray}
For momentum independent gauge fixing parameter $\xi$, the mWI in the presence
of the cutoff term \eq{eq:A-cutoff} reads
\begin{eqnarray} 
\del_\alpha \Gamma_{k,T}[A] =
T\sum_n\int \frac{d^3 q}{(2\pi)^3}\,\tr\,\left[
\frac{1}{n^2\xi}\,n_\mu\partial_\mu \alpha\ n_\nu A_\nu
+\frac{1}{2} [\alpha, R_k^{\mu\nu}]\, G_{k,T}{}_{\nu\mu}[A]\right]\,.
\label{eq:mWI} 
\end{eqnarray} 
The two terms on the r.h.s.~are remnants from the gauge fixing and the
coarse-graining, respectively.  The mWI \eq{eq:mWI} is compatible with
\eq{eq:flow}
\cite{D'Attanasio:1996jd,Litim:1998qi,Litim:2002ce,%
Freire:2000bq,Pawlowski:2005xe},
i.e.\ a solution to \eq{eq:mWI} at some scale $k=\Lambda$ remains a solution
for $k<\Lambda$ if $\Gamma_{k,T}[\phi]$ is integrated according to the flow
equation. In particular, the terms proportional to $R_k$ vanish for $k\to 0$,
thereby ensuring gauge invariance for physical Green functions.  The mWI
related to the thermal flow \eq{eq:Tgaugeflow} follows from \eq{eq:mWI} as
\begin{eqnarray}
&&\hspace{-1.5cm}
\del_\alpha\bar\Gamma_{k,T}[A]=
\0g2 \int \0{d^3 q}{(2\pi)^3}\,\tr\,  \left[ 
\int \frac{dq_0}{2\pi} [\alpha, R_k] \, G_{k,0}[A]
-T\sum_n[\alpha, R_k] \, G_{k,T}[A] \right]
\label{delmWI} 
\end{eqnarray} 
The compatibility of \eq{delmWI} with \eq{eq:Tgaugeflow} is a direct
consequence of the compatibility of \eq{eq:mWI} with \eq{eq:flow}.  The linear
term related to the gauge fixing [the first term in \eq{eq:mWI}] has cancelled
out, since, as emphasized in Sect.~\ref{QvsT}, we are only looking at fields
$\phi$ at temperature $T$ and corresponding gauge transformations. This also
implies -- up to modifications for topologically non-trivial configurations --
that $\alpha$ has to be periodic. Apart from this simplification, the same
reasoning as for \eq{eq:mWI} above applies.\step

With a mass-like regulator, however, we can go a step further. For a regulator
as in \eq{R-mass}, the right hand side in \eq{delmWI} vanishes since then
$[\alpha,R_k]=0$. All coarse-graining dependence of \eq{delmWI} drops out for
arbitrary scale $k$, and not only in the limit $k\to 0$.  This is an immediate
consequence of $R_k$ being momentum independent and the axial gauge fixing
\cite{Litim:1998qi,Litim:2002ce,Simionato:1998iz}, and reduces \eq{eq:mWI} to
the standard Ward Identity in the presence of an axial gauge fixing.  It
follows, that \bm{c}\di \del_\alpha\bar\Gamma_{k,T}[A]=0
\label{gaugeinv} 
\eg and we end up with the statement that \eq{eq:Tflowmass} corresponds to
a gauge invariant thermal Wilsonian renormalisation group for
$\bar\Gamma_{k,T}[\phi]$, valid at any scale $k$.\step

The above flow can be used for computing the self-energy and the thermal
pressure for one-loop resummed perturbation theory. Following the computations
in Sect.~\ref{sec:resum} and \ref{sec:deriv} we arrive at the well-known
result for the thermal mass
\begin{eqnarray}\label{eq:selfen}
m_{\rm gluon}^2=\016 g^2 T^2 N\,.  
\end{eqnarray}  
The thermal mass \eq{eq:selfen} can be re-inserted into the flow.
However, full resummations require also the one-loop running of
propagator and classical vertices. This is straightforward but
tedious, and the results shall be presented elsewhere. Finally we want
to comment on the computation of the thermal pressure.  It is
well-known that the computation requires the correct normalisation of
the path integral, only physical degrees of freedom contribute to the
pressure. In the present formalism this is the correct choice of $\bar
C_{k,T}$. This amounts to the projection of the field independent part
of the flow in momentum space onto the transversal parts of momenta
orthogonal to the gauge fixing vector $n$. Such a procedure
immediately results in
\begin{eqnarray}\label{eq:SUNpress} 
P=(N^2-1)\frac{\pi^2T^4}{45}\,.  
\end{eqnarray} 
The computation of higher contributions, in particular the
resummation, requires the full one loop running of vertices and
propagators. We also emphasise that \eq{eq:3/2} already provides the
closed expression for the resummation. We hope to report on this
matter in near future.

\section
{Discussion and conclusions}\label{Discussion}

In this paper, we have developed a thermal renormalisation group in
terms of the functional
$\bar\Gamma_T[\phi]=\Gamma_T[\phi]-\Gamma_0[\phi]$. The approach
implements a thermal projection at every integration step of the
Wilsonian flow. From $\bar\Gamma_T[\phi]$ all thermal observables --
including the thermal pressure and the thermal self-energy -- can be
obtained.  Our flow for $\bar\Gamma_{T}$ is free of infrared and
ultraviolet divergences and therefore well-suited for numerical and
analytical studies. The running zero-temperature effective action
serves as a boundary condition for the flow. The flow for
$\bar\Gamma_T[\phi]$ can be solved with standard techniques including
expansions in derivatives or vertex functions, and is not confined to
the weakly coupled domain. Moreover, the thermal flow derives from a
path integral representation of the theory.  This ensures that no
double counting occurs in the flow for $\bar\Gamma_T[\phi]$ and
systematic solutions thereof.  While the present construction is based
on the imaginary time formalism, it is straightforward to implement
these ideas even in a real-time formulation.  \step

For scalar theories, we have shown that the leading-order resummed
perturbation theory follows independently of the scheme, and independently of
the truncation. Quantitatively, this has been worked out to leading order
within a thermal derivative expansion.  Furthermore, an optimised thermal flow
for the effective potential has been provided. Its simple analytic form allows
for stable numerical integrations, in particular at strong coupling. As an
aside, we point out how \cite{Drummond:1996ap} -- where an infrared regulated
resummation for the thermal pressure of scalar theories has been proposed --
is related to our work.  If we restrict the full thermal flow
\eq{eq:Tflowmass} to scalar fields, and to leading order in the derivative
expansion of the effective action at vanishing field, then \eq{eq:Tflowmass}
corresponds to the imaginary-time analogue of \cite{Drummond:1996ap}. Hence,
\cite{Drummond:1996ap} has the interpretation of a leading-order Wilsonian
flow with mass term cutoff. Since \eq{eq:Tflowmass} defines the
field-dependent effective action, it allows for an extension of
\cite{Drummond:1996ap} to higher order operators as well as to fermions and
gauge fields.\step

For gauge theories, the following picture has emerged: the standard Wilsonian
flow equation \eq{eq:flow}, equipped with the modified Ward identity
\eq{eq:mWI}, allows for the consistent computation of infrared quantities
starting with an initial action in the ultraviolet. Gauge invariance is
ensured in the physical limit. A gauge invariant implementation for all scales
$k$ -- in the Wilsonian sense -- is problematic due to the poor performance of
a mass-term regulator for quantum fluctuations in the ultraviolet. The
important new result is that the thermal flow \eq{eq:Tflow}, instead, stays
well-defined even for a mass-like regulator \eq{eq:Tflowmass} since the zero
temperature contribution renders the flow finite.  A mass term regulator fully
pays off when combined with the axial gauge fixing, as it implies a
gauge-invariant thermal flow for all scales $k$. The difference to the flow
\eq{eq:flow} stems now from the initial condition, which is no longer the bare
action, but the running $T=0$ quantum effective action, or some approximation
to it.\\[3ex]

{\it Acknowledgements:}\ DFL is supported by an EPSRC Advanced
Fellowship. JMP is supported by the DFG under contract GI328/1-2.


\begin{thebibliography}{99}
\def\BOOK#1#2#3#4{#1 {\it #2}, #3, #4}
\def\PRA#1#2#3#4#5{ #1   Whys.~Rev.~{\bf A #3} (19#4) #5}
\def\PRB#1#2#3#4#5{#1   Phys. Rev.~{\bf B #3} (19#4) #5}
\def\PRL#1#2#3#4#5{#1   Phys. Rev.~Lett.~{\bf #3} (19#4) #5}
\def\PRC#1#2#3#4#5{#1   Phys. Rev.~{\bf C #3}  (19#4) #5}
\def\PRD#1#2#3#4#5{#1   Phys. Rev.~{\bf D #3} (19#4) #5}
\def\PRE#1#2#3#4#5{#1   Phys. Rev.~{\bf E #3} (19#4) #5}
\def\PRep#1#2#3#4#5{#1   Phys. Rep.~{\bf  #3} (19#4) #5}
\def\NPB#1#2#3#4#5{#1   Nucl. Phys.~{\bf B #3} (19#4) #5}
\def\PLB#1#2#3#4#5{#1   Phys. Lett.~{\bf B #3} (19#4) #5}
\def\ibid#1#2#3#4#5{#1   {\it ibid.~}{\bf #3} (19#4) #5}
\def\PTP#1#2#3#4#5{#1   Prog. Theor.~Phys.~{\bf B #3} (19#4) #5}
\def\SSC#1#2#3#4#5{#1   Solid State Comm.~{\bf  #3} (19#4) #5}
\def\EPL#1#2#3#4#5{#1   Europhys. Lett.~{\bf #3} (19#4) #5}
\def\JCP#1#2#3#4#5{#1   J.~Phys. (Paris) {\bf  #3} (19#4) #5}
\def\JPA#1#2#3#4#5{#1   J.~Phys. {\bf A  #3} (19#4) #5}
\def\JPB#1#2#3#4#5{#1   J.~Phys. {\bf B  #3} (19#4) #5}
\def\JPC#1#2#3#4#5{#1   J.~Phys. {\bf C  #3} (19#4) #5}
\def\ZPC#1#2#3#4#5{#1   Z.~Phys. {\bf C  #3} (19#4) #5}
\def\JETP#1#2#3#4#5{#1   Soviet Physics JETP Lett.~{\bf #3} (19#4) #5}
\def\MPLA#1#2#3#4#5{#1   Mod.~Phys. Lett.~{\bf A  #3} (19#4) #5}
\def\PA#1#2#3#4#5{#1   Physica {\bf A  #3} (19#4) #5}
\def\PS#1#2#3#4#5{#1   Physics {\bf   #3} (19#4) #5}
\def\AP#1#2#3#4#5{#1   Ann. Phys. {\bf  #3} (19#4) #5}
\def\IJMPA#1#2#3#4#5{#1   Int.~J. Mod. Phys.~ {\bf A  #3} (19#4) #5}
\def\LNC#1#2#3#4#5{#1   Lett.~Nuevo Cimento {\bf   #3} (19#4) #5}
\def\PPR#1#2{#1 {\tt #2}}
\def\and#1#2#3{{\bf #1} (19#2) #3}


\bibitem{StrongQGP}
U.~Heinz, nucl-th/0412094.

\bibitem{Kraemmer:2003gd}
  U.~Kraemmer and A.~Rebhan,
  Rept.\ Prog.\ Phys.\  {\bf 67} (2004) 351
  [hep-ph/0310337].

\bibitem{Andersen:2004fp}
  J.~O.~Andersen and M.~Strickland,
  Annals Phys.\  {\bf 317} (2005) 281
  [hep-ph/0404164].

\bibitem{Blaizot:2003iq}
  J.~P.~Blaizot, E.~Iancu and A.~Rebhan,
  Phys.\ Rev.\ D {\bf 68} (2003) 025011
  [hep-ph/0303045].

\bibitem{Morris:1998da}
  T.~R.~Morris,
  Prog.\ Theor.\ Phys.\ Suppl.\  {\bf 131} (1998) 395
  [hep-th/9802039].

\bibitem{Litim:1998yn} D.~F.~Litim, {\it Wilsonian flow equation and thermal
    field theory}, refereed contribution in: U.~Heinz (Ed.), {\it Thermal
    Field Theories and their Applications}, [hep-ph/9811272].

\bibitem{Litim:1998nf} D.~F.~Litim and J.~M.~Pawlowski, in {\it The Exact
  Renormalization Group}, Eds.~Krasnitz et al, World Sci (1999) 168 
  %
  [hep-th/9901063].


\bibitem{Berges:2000ew}
  J.~Berges, N.~Tetradis and C.~Wetterich,
  %
  Phys.\ Rept.\  {\bf 363} (2002) 223.
\bibitem{Polonyi:2001se}
  J.~Polonyi,
  %
  Central Eur.\ J.\ Phys.\  {\bf 1} (2004) 1
  [hep-th/0110026].

\bibitem{Salmhofer:2001tr}
  M.~Salmhofer and C.~Honerkamp,
  %
  Prog.\ Theor.\ Phys.\  {\bf 105} (2001) 1.

\bibitem{Pawlowski:2005xe}
  J.~M.~Pawlowski,
  hep-th/0512261.


\bibitem{Tetradis:1992xd}
  N.~Tetradis and C.~Wetterich,
  Nucl.\ Phys.\ B {\bf 398} (1993) 659.

\bibitem{Liao:1995gt}
  S.~B.~Liao and M.~Strickland,
  Phys.\ Rev.\ D {\bf 52} (1995) 3653
  [hep-th/9501137].

\bibitem{D'Attanasio:1996zt}
  M.~D'Attanasio and M.~Pietroni,
  Nucl.\ Phys.\ B {\bf 472}, 711 (1996)
  [hep-ph/9601375].



\bibitem{D'Attanasio:1996fy}
  M.~D'Attanasio and M.~Pietroni,
  Nucl.\ Phys.\ B {\bf 498} (1997) 443
  [hep-th/9611038].

\bibitem{Comelli:1997ru}
  D.~Comelli and M.~Pietroni,
  Phys.\ Lett.\ B {\bf 417} (1998) 337
  [hep-ph/9708489].

\bibitem{Bergerhoff:1998xx}
  B.~Bergerhoff and J.~Reingruber,
  Phys.\ Rev.\ D {\bf 60} (1999) 105036
  [hep-ph/9809251].

\bibitem{Andersen:1999dy}
  J.~O.~Andersen and M.~Strickland,
  Phys.\ Rev.\ A {\bf 60} (1999) 1442
  [cond-mat/9811096].

\bibitem{Schaefer:1999em}
  B.~J.~Schaefer and H.~J.~Pirner,
  Nucl.\ Phys.\ A {\bf 660} (1999) 439
  [nucl-th/9903003].

\bibitem{Freire:2000sx}
  F.~Freire and D.~F.~Litim,
  Phys.\ Rev.\ D {\bf 64} (2001) 045014
  [hep-ph/0002153].

\bibitem{Tflow2003}
C.~Honerkamp and M.~Salmhofer, cond-mat/0105218.

\bibitem{Baier:2003ex}
  T.~Baier, E.~Bick and C.~Wetterich,
  cond-mat/0309715.


\bibitem{Braun:2003ii}
  J.~Braun, K.~Schwenzer and H.~J.~Pirner,
  Phys.\ Rev.\ D {\bf 70} (2004) 085016
  [hep-ph/0312277].

\bibitem{Blaizot:2004qa}
  J.~P.~Blaizot, R.~Mendez Galain and N.~Wschebor,
  cond-mat/0412481.

\bibitem{Braun:2006jd}
  J.~Braun and H.~Gies,
  JHEP {\bf 0606}, 024 (2006)
  [hep-ph/0602226];
  hep-ph/0512085.

\bibitem{Schaefer:2006ds}
  B.~J.~Schaefer and J.~Wambach,
  hep-ph/0603256.

\bibitem{Litim:1998qi}
  D.~F.~Litim and J.~M.~Pawlowski,
  Phys.\ Lett.\ B {\bf 435} (1998) 181
  [hep-th/9802064].

\bibitem{Litim:2002ce}
  D.~F.~Litim and J.~M.~Pawlowski,
  JHEP {\bf 0209} (2002) 049
  [hep-th/0203005].



\bibitem{Simionato:1998iz}
  M.~Simionato,
  Int.\ J.\ Mod.\ Phys.\ A {\bf 15} (2000) 2153
  [hep-th/9810117].

\bibitem{Simionato:2000ut}
  M.~Simionato,
  Int.\ J.\ Mod.\ Phys.\ A {\bf 15} (2000) 4811
  [hep-th/0005083].

\bibitem{Panza:2000tg}
  A.~Panza and R.~Soldati,
  Phys.\ Lett.\ B {\bf 493} (2000) 197
  [hep-th/0006170].

\bibitem{Litim:2000ci}
  D.~F.~Litim,
  %
  Phys.\ Lett.\ B {\bf 486} (2000) 92 [hep-th/0005245];
  %
  Nucl.\ Phys.\ B {\bf 631} (2002) 128 [hep-th/0203006].

\bibitem{Litim:2005us}
  D.~F.~Litim,
  JHEP {\bf 0507} (2005) 005
 [hep-th/0503096].


\bibitem{Litim:2001up}
  D.~F.~Litim,
  %
  Phys.\ Rev.\ D {\bf 64} (2001) 105007
  [hep-th/0103195].



\bibitem{Litim:2002xm}
  D.~F.~Litim and J.~M.~Pawlowski,
  Phys.\ Rev.\ D {\bf 66} (2002) 025030
  [hep-th/0202188];
  Phys.\ Lett.\ B {\bf 546} (2002) 279
  [hep-th/0208216].


\bibitem{D'Attanasio:1996jd}
  M.~D'Attanasio and T.~R.~Morris,
  Phys.\ Lett.\ B {\bf 378}, 213 (1996)
  [hep-th/9602156].

\bibitem{Freire:2000bq}
  F.~Freire, D.~F.~Litim and J.~M.~Pawlowski,
  Phys.\ Lett.\ B {\bf 495} (2000) 256
  [hep-th/0009110].


\bibitem{Drummond:1996ap}
  I.~T.~Drummond, R.~R.~Horgan, P.~V.~Landshoff and A.~Rebhan,
  Phys.\ Lett.\ B {\bf 398} (1997) 326
  [hep-th/9610189].


\end{thebibliography}
\end{document}